# Some Aspects of a Software Reliability Problem

Anton Petrov[1], Elena Popova[1], Alexander Petrov[2]

[1] KubSAU, Russia
[2] AGH, Krakow, Poland
anton.a.petrov@gmail.com

**Abstract.** Obviously, the dynamism of software reliability research has speeded up significantly in the last period, and we can state the fact that its intensity is approaching, and in some cases is ahead of the information systems hardware reliability research intensity. Reliability of software is much more important than its other characteristics, such as runtime, and although the absolute reliability of modern software is apparently unattainable, there is still no generally accepted measure of reliability of computer programs. The article analyzes the reasons for the situation and offers an approach to solving the problem. The article touches upon the issue of general characteristics of information systems software life cycle. Considered software application reliability questions and use of fail-safe ensuring programming. Also presented basic types of so-called virus programs that lead to abnormal functioning of information systems. Much attention is given to presenting some known models used for software debugging and operating. So, this review paper consists of four sections: information systems software process creation, reliability of information systems software, using of fail-safe programs and estimation of software reliability according the results of adjusting and normal operation.

## 1 Information Systems Software Development Process

### 1.1 Problem Overview

To ensure the reliability of programs, many approaches have been proposed, including organizational development methods, various technologies and technological software tools, which require significant resources. However, the lack of universally accepted criteria for reliability does not allow us to answer the question of how much more reliable the software becomes in compliance with the proposed procedures and technologies. Thus, the priority of the task of assessing reliability should be higher than the priority of the task of ensuring it, which is not observed.

Analyzing existing publications, we can conclude – the issue of ensuring the reliability of programs is considered more important than the question of its evaluation. The situation looks paradoxical: it is obvious that before improving some characteristic, one must learn to measure it, it is also necessary to have a unit of measurement. The main reason for this situation is rooted in the fact that the source of unreliability of the programs is the errors contained in them, and if there are no errors, then the program is reliable. Essentially, all measures to ensure the reliability of programs are aimed at minimizing (if not eliminating at all) errors during development and at the earliest possible time to identify and eliminate them after the program is made. It should be noted that error-free programs, of course, exist, but modern software systems are too large and almost inevitably contain errors. Although this circumstance is noted by many authors and is known to any practical programmer, there seems to be a psychological barrier that does not allow recognizing the fact that errors in software are inevitable reality. Since there is no exact criterion for determining the maximum size of an error-free program, there is always the hope that they will not remain in this software system.

There is another psychological circumstance. As you know, the issue of reliability for the equipment is well developed. The source of unreliability of the equipment is objective factors that are not subject to man (power surges, alpha particles, etc.), therefore, humanity has long come to terms with the idea that absolutely reliable equipment does not exist, and we can only talk about the degree of reliability expressed in some units (for example, the average time between two consecutive failures). The source of insecurity is the error programs that people who create and use them make, so it seems that the only problem is to make (or teach) them work correctly.

The reason is that the problem of choosing the unit of measurement for the reliability of a computer program cannot be solved within the framework of the industrial approach, which currently occupies an increasingly dominant position in programming. The most typical example is the use, by analogy with equipment, as a measure of the reliability of the average time program between two consecutive erroneous operations.

The analogy method, of course, is universal, but we should not forget that any analogy has limits of applicability. In this case, since we are talking about a fundamental concept (unit of measurement), you should not just transfer the reliability characteristics of the equipment to the programs but use more fundamental analogies. First, it is useful to recall where the reliability characteristics of the equipment come from. Reliability, in the final analysis, is a statistical concept, i.e. it is assumed that there is a certain (sufficiently large) number of identical samples, tests, etc., there is also an element of randomness. The study of random phenomena is devoted to a special section of mathematics: probability theory. The basic concept of this theory is the space of elementary events (sample space, space of outcomes), on which a certain (probabilistic) measure is specified. The random variable, according to the theory, is a function defined on the space of elementary events. Finally, as a measure of reliability, some characteristics of a random variable are used (as a rule, mathematical expectation).

Thus, a consistent probabilistic approach to the study of reliability consists in analyzing the object under study (aircraft, security systems, computer programs, etc.), constructing, based on "physical" considerations about its nature, spaces of elementary events, introducing a probability measure on them and consideration of random variables.

Unfortunately, the first stage of research - the analysis of an object and the construction of spaces of elementary events - is usually omitted and immediately proceed to the consideration of random variables, losing sight of the fact that a random variable is actually a function defined on the space of elementary events.

Before talking about the reliability of the object, it should be clarified what is meant by the object. As you know, a computer program has several different forms (or representations): external specifications, source code, executable code, etc. The generally accepted point of view is that a program is an object that is invariant with respect to the forms of its representation. According to this point of view, external specifications, source codes in languages of different levels, as well as executable codes for different processors, have different forms of representing the same program. This point of view is useful in software development, because it allows you to identify the most essential for the application program properties that are common to all its representations, but it is unproductive if, for example, we are talking about such a quantitative characteristic as execution time: it is clear that this characteristic refers only to one of the forms of representation - the executable code and, in addition, depends not only on the program, but also on the type of processor.

At an intuitive level, the concept of program reliability reflects the fact that it cannot always give the correct result. This means that the reliability of a program is a characteristic of its executable code. The executable code corresponds to the source text in the same way as, for example, the electric motor and its drawings: we can talk about the reliability of the manufactured product, but it makes no sense to talk about the reliability of the description, drawing, text. Two functionally identical programs written in different languages or prepared for different types of machines or for the same machine, but using different compilers, should be considered different in terms of reliability. A program is considered correct if it does not contain errors. Such a program does not give incorrect results, i.e. she is reliable. This fact gave rise to a false idea that the number of errors in the program can be considered the most natural measure of reliability [1]. Quite a lot of work has been done, in which various methods were proposed for estimating the number of errors remaining in the program according to the results of its testing, including the method of "clogging" with known errors, however, as the considerations below show, the number of errors in the program has nothing to do with it reliability: the number of errors in the program is the "unobservable" value; it is not the errors themselves that are observed, but the result of their manifestation.

Incorrect operation of the program may be the result of not one, but several errors at once.

Errors can compensate each other, so that after fixing a single error, the program may start to "work worse." Reliability characterizes the frequency of occurrence of errors, but not their number; at the same time, it is well known that errors occur at different frequencies: some errors remain undetected after many months and even years of operation, but, on the other hand, it is not difficult to give examples where one single error leads to incorrect operation of the program for any initial data , i.e. to zero reliability. It should also be noted that if the number of errors is considered as a measure of reliability, then in the terminology of probability theory this number is a random variable, but the main question - in which space of elementary events it is assigned - was not addressed anywhere.

Finally, it is important to emphasize that, from the point of view of reliability, as a result of error correction or any other correction, a new program with a different reliability indicator than before correction is obtained. Thus, the number of errors in the program characterizes rather than the program, but its manufacturers and the tools used [10].

## 1.2 The life cycle of a software

The life cycle of a software (product) begins with the determining of its technical project development and ends at termination of its use [1].

There are following processes that executed during the life cycle of the software: five basic processes; eight supporting and four organizational processes.

Basic processes of life cycle structure consist of five processes that serve to the major constituents during the software life cycle.

The major constituencies are the subjects that initiate, develop, operate and conduct software maintenance. The main participants are customer, supplier, developer, operator and software maintenance supporter. The main procedures include the following steps:

- order – specifies the actions of the customer organization which order system, software product or software service,
- supply – establishes the supplier organization actions that provide the system software product or the software service,
- development – defines the actions of developer's organization that identifies and designs the software,
- operation – sets the actions of operator-organization that provides the (automated) services of information system in its current state to the users,
- support – determines the action of maintainer organization which provides services in support of the software, therefore manages modification of the software in order to maintain it in proper and operational condition. These processes involve the transfer and removal of software.

The support is considered as an integral part of the process that attends the latter; it has a clearly defined purpose and contributes to the successful software quality implementation. The process of support is applied and implemented in the proceedings according to its requirements [2]. The structure of the software life cycle support process includes the following proceedings:

- documentation – defines the steps to register the information, which was obtained during the life cycle course,
- configuration – defines actions regarding the software configuration management,
- quality assurance – determines the actions to acquire objective assurance that the software products and processes are ready to meet the specified requirements and adhere to its established plans.

As a method of quality assurance processes, it can be used shared inspection, auditing, verification and validation:

- Verification – defines the actions of the customer, supplier or independent participant about verification software with varying degrees of depth, depending on the software features.

- Validation – defines the actions of the customer, supplier or independent participant regarding the obtained in the frameworks of programming project; software validation conducting.

- Shared scrutiny – determines the actions to assess the status and results of a definite action. This process can be applied by any two members, one of which (the party that review) evaluates actions of another party (party whose actions are reviewed) in a mutual discussion.

- Audit – defines actions to determine compliance with requirements, plans and contract. This process can be applied by any two members, one of which (the participant which checks) conducts an audit of the software or the actions of another participant (the participant which is being checked).

- Problem solving – identifies actions for analysis and removal of problems (including non-compliance), of any category of their nature and causes that were identified during the development, operation, maintenance or during other processes performing.

Organizational processes regarding the software lifecycle are used by an organization on order to establish and implement the basic structure which consists of life cycle interconnected processes and pertinent personnel, as well as for structures and processes continuous improvement. Their implementation is usually beyond the scope of improvement of specific projects and contracts, but the experience gained in projects and contracts can be used for organization performance improvement. Organizational processes include processes:

- controls – defines the actions regarding control, including project management during life cycle,

- infrastructure making – defines the essential steps for the basic structure of the life cycle processes building,

- foundation – defines the basic steps which an organization (that may be a customer, supplier, developer, operator, developer or manager of a process), conducts in order to create, measure, control and improve the processes of life cycle that it supports,

- training – determines the actions to provide appropriate staff schooling.

## 2  Reliability of information systems software

Software reliability is the ability of a software product to fail to perform certain functions under specified conditions for a given period of time with a high probability. The degree of reliability is characterized by the probability of the software product working without failure for a certain period of time. Programs for modern information systems can accrue significant number (hundreds, thousands, tens or hundreds of thousands) of simple commands. For many reasons at writing programs it may be revealed the errors. In the environ of programmer's humor, they say, there is no software without bugs, but there are programs with errors not found. Blunders are detected at the stage of working out the programs, but to check the program totally according all possible modes is not usually viable, as result there is no conviction that all errors are found [2-4]. There used a statistical approach to the process of errors detecting during the programming analysis.

Such process can be characterized according the function:

$$\frac{f(t)}{R} \tag{1}$$

where $f(t)$ – number of identified and corrected errors per time unit in the program, which has $R$ – number of commands.

$$\frac{f(t)}{R} = \frac{d\varepsilon_n}{dt} = \frac{\varepsilon_n(t+\Delta t) - \varepsilon_n(t)}{\Delta t} \tag{2}$$

where $\varepsilon_n(t)$ – number of identified and corrected errors during time $t$ per one command.

Accordingly:

$$\varepsilon_n(t) = \frac{1}{R}\int_0^t f(t)dt \tag{3}$$

Function $f(t)$ can be determined experimentally during pilot testing of program by means of fixing the detected errors number. Problem of $f(t)$ definition is simplified if:

$$f(t) = \frac{\varepsilon_0}{\tau_0} e^{-\frac{t}{\tau_0}} \tag{4}$$

where $\varepsilon_0$ and $\tau_0$ are parameters $f(t)$ which are determined during working out.

Then:

$$\varepsilon_n(\tau) = \frac{1}{R}\int_0^\tau f(t)dt = \frac{\varepsilon_0}{R}(1 - e^{-\frac{\tau}{\tau_0}}) \tag{5}$$

At $\tau \to \infty$ $\varepsilon_n(\infty) = \frac{\varepsilon_0}{R}$ or $\varepsilon_0 = R\varepsilon_n(\infty)$. From there it follows that $\varepsilon_0$ is the total number of errors in the program prior to testing. Since the testing process cannot subsist too lengthy then in the program always will remain some errors:

$$\varepsilon(\tau) = \frac{\varepsilon_0}{K} - \varepsilon_n = \frac{\varepsilon_0}{R} e^{-\frac{\tau}{\tau_0}} \tag{6}$$

where $\varepsilon(\tau)$ – the number of errors found per one command. If to anticipate that errors are uniformly distributed throughout the program, the occurrence probability of errors $P(t)$ during $\Delta t$ will be proportional to the tempo pace $\delta$ of information system (that is the average number of commands changes per time unit) and to the number of errors left in the program, that is:

$$P(\tau) = \varepsilon(\tau)\delta\Delta t \qquad (7)$$

Pursuing the analogy between the process of errors and failures of objects ($P(t) = \lambda \Delta t$) it may be concluded that the intensity of errors $\varepsilon(\tau)$ does not depend on time $t$ but is determined only by the interval $\Delta t$ at which it is estimated the probability of error. From there operation time till "failure", which is due to the error emergence in the program, will make:

$$T(\tau) = \frac{1}{\varepsilon(\tau)\delta} = \frac{R}{\varepsilon_0 \delta} e^{\frac{\tau}{\tau_0}} \qquad (8)$$

Analysis of $T(\tau)$ changes can serve as a basis for the program working-out operation timing $\tau$ choice, namely, the testing ends when the value $T(\tau)$ becomes large enough.

In case when it is possible to estimate material losses $C_n$ via occurrence of errors in the calculations, then the operation time testing $\tau$ can be quantified as the following. If during time $T_p$ of program operation, it fails $\frac{T_p}{T(\tau)}$ times, it will cause a total loss $C_n \frac{T_p}{T(\tau)}$. The process of program testing requires some time-consuming computations and other costs associated with it. If to mark $C_0$ the cost of a single time testing, then during time $\tau$ such costs will amount $C_0 \tau$. Accordingly, the total loss $C$ because of errors and costs for working testing of programs will be:

$$C = \frac{C_n T_p}{T(\tau)} + C_0 \tau = \frac{C_n T_p \varepsilon_0 \delta}{R} e^{-\frac{\tau}{\tau_0}} + C_0 \tau \qquad (9)$$

From there:

$$\frac{dC}{d\tau} = \frac{-C_n N_p \varepsilon_0 \delta}{R \tau_0} e^{-\frac{\tau_M}{\tau_0}} + C_0 = 0$$

or

$$\tau_M = -\tau_0 \ln \frac{C_0 R \tau_0}{C_n T_p \varepsilon_0 \delta},$$

where $\tau_M$ – time of working out which will provide the minimum $C$.

In this case, when it is required to eliminate an error in the program it is advisable to use the "backup". Here particular problem is solved by several programs, each of them is developed by independent teams of programmers and in its basis lays various algorithms, furthermore, the results of programs computations are compared, and they are considered true if they match. Since the errors emergence in software is an improbable event, the occurrence of two or more of such events is practically impossible.

## 3    Fault Tolerant Information Systems Software

Fail-safe programs are designed usually by frequent repetition of calculations at the levels of micro-operations, operations, commands, program modules or the entire program.

To improve reliability against the failures of the entire information system it is widely used method of repeated execution of programs at the level of program module. Its essence is that the program is split into several modules, each of them is executed twice, and the results are compared. If the results of the first and second calculations coincide, it is considered that the results obtained are true and then it may to proceed to the next step of (operational) calculations. At disagreement, the computation is repeated until the two received results will be the same. The substantial advantage of this method is its simplicity. At drawing up the program it is required only to provide the appropriate actions, the method does not require additional costs for hardware. The disadvantage of this method is in the time for problems solving more than twice growth, and in inability to detect errors caused by the failures.

Performance of information system at using the method of double execution depends on the number of modules into which is to be split the program. Indeed, the greater length of the modules determines also the large probability of failures. So, instead of two, it will be required to repeat computations three or more times, which will increase the problem solution time. On the other hand, at small length of modules, most of time will be spent on comparing and recording the calculations results executed within individual program modules into the memory devices.

In this regard, there emerges a problem of finding an optimal number of modules into which a program should be split, namely in such way that the time $T_P$ for problem solving will be minimal. We introduce the notation: $T$ – time for solving of problem at a single execution of the program; $t$ – duration of calculation at a single module; $p(t)$ – the probability of no failure during time $t$. Then the ratio $\frac{T}{t}$ will determine the number of modules onto which should be divided a program. We can determine the probabilities of two-, three-, or even $i$ times execution repetition for any program module. If the failures are independent events, then the probability that a given program module will be executed twice will be equal to the probability of no failure at the first and second executions, so that:

$$p_2(t) = p_1^2(t) \qquad (10)$$

Subsequently, the probability $p_1(t)$ at fixed $t$ will be denoted as $p_1$.

Similarly, $g$ is the probability that in one of the two preceding calculations has occurred failure, but at the third computation was obtained correct result, that is:

$$p_3 = 2p_1^2(1-p_1) = 2p_1^2 \qquad (11)$$

where $q = 1-p$. In general, $p_3$ equals to the probability that at the $i$-st and in one of the preceding calculations the failures were absent, and in the others, already passed, the failures were presented, that is:

$$p_3 = (i-1)p_1^2 q^{i-2} \qquad (12)$$

Thus, the average number of computing will be equal:

$$A = \sum_{i=2}^{\infty} p_i = \sum_{i=2}^{\infty} i(i-1) p_1^2 q^{i-2} \qquad (13)$$

It is easy to show that $\frac{A}{p_1^2} = \frac{2}{(1-q)^3}$. Hence, we have $A = \frac{2}{p_1}$. Thus, the time spent on calculating will amount $\frac{2T}{p_1}$. Time $T_3$ which is required to conduct comparisons and to write intermediate calculations into the memory device, depends on the type of memory storage device used, on the number of intermediate results $k$ and on the number of steps $\frac{T}{t}$ of the program, that is:

$$T_3 = \frac{T}{t} f\left(k, \frac{T}{t}\right) \qquad (14)$$

where $f\left(k, \frac{T}{t}\right)$ – the average time of comparisons operations and of the recourse to the memory device for one module of program results recording. If we assume that $f\left(k, \frac{T}{t}\right) = const = a$ then:

$$T_p = \frac{2T}{p_1(t)} + \frac{Ta}{t} = T\left(\frac{2}{p_1 t} + \frac{a}{t}\right) \qquad (15)$$

For some types of information systems it was experimentally found that:

$$p(t) = e^{-\lambda \cdot t} \qquad (16)$$

where $\lambda$ – the intensity of failures. In this case, $T_P$ accepts the minimum value for $t$, which can be determined from the equation:

$$\frac{dT_P}{dt} = 2\lambda e^{\lambda \cdot t} - \frac{a}{t^2} = 0 \qquad (17)$$

Thus, the value of $T_P$ can determine the optimal length of the program section and corresponding to it number $\dfrac{T}{t}$ of stations at which $T_P$ will be minimal.

The cause of incorrect functioning of the information system may be also a presence in its so-called virus software programs designed to insert undue distortions into computations, deleting files and creating conditions for the abnormal functioning of information system.

In accordance to The Codifier of Information System Crimes of the General Secretariat of Interpol, viruses are classified to QD – the data information systems changes, within which they are classified as following:
- QDL – logic bomb,
- QDT – trojan horse,
- QDV – virus of information system,
- QDW – worm of information system,
- QDZ – other's data changes.

Logic bomb – secretly inserts into a program a set of commands that should work only once but starting at definite circumstances.

Trojan horse – provides an introduction into someone else's program of such commands, that allow conducting some foreign, not planned by the proprietor of program operations, while at the same time they preserve the general performance of the host program liability.

Virus software in information system – a specially written programs that can "attribute" themselves to other programs (that is to "infect" them), to reproduce and give birth to new viruses to conduct various undesirable actions in the information system.

Worm software in information system – special self-distributing software that makes editing data or programs of information system, without legal right, by transfer, introduction or spread through a network of information systems.

The share of errors or lockups of information system caused by viruses is about 10% to 30%. There are known more than 10,000 viruses and about 100 antivirus programs designed to combat them. There exist viruses (self-instructed, polymorphic, macro viruses etc.) that can counteract the antiviral programs. One of such virus varieties implements "settlement" in the anti-viral program. Usually an antivirus program gives a signal of its infection if such event took place. Time required to cure a virus is on average from 15 to 30 minutes. The most dangerous virus is a virus that settles in the executive file. Most viruses "are working" apparently correctly and do not cause information system deadlocks. But among them are also those which completely erase the hard disk system areas or subdirectories of information files. In 90% of cases the viruses penetrate information systems through the network. Normally local networks themselves do not distribute viruses. But users who work with memory devices which are damaged with viruses deliver to such network a lot of trouble.

The symptoms of information system infection with a virus are:
- increase of number of errors and lockups of information system,
- slow down the programs loading,
- problems (various slowdown and errors) with printer operation,

- drive lights flashing when it does not have to read/write,
- resizing the volume of executable programs, reducing the major available memory.

The volume shortest are destroying viruses; their size does not exceed 20 kilobytes. Most viruses have volumes up to 100 kilobytes or more. Recently a lot of trouble to the users deliver macro viruses.

Quality of antiviral program is defined according the following characteristics listed in descending order of importance:
- reliability and ease of operation (no technical problems, does not require a user's special training),
- number of all type viruses finding, an ability to scan the documents/spreadsheets files, packed files and an ability of contaminated objects curing,
- speed of operation and a variety of another practical features.

If a user has several effective antiviral programs and utilizes it, the most reliable protection against viruses is in its prevention as:
- creating of regular backups (for example, once a week – complete, every day – partial copy); the presence of uninfected copies allows rewriting a "sick" file, the presence of infected but not damaged copies will allow to restore files after the virus removal;
- making backup copies of installation memory media before the installing of new software (if they are installed on the infected program the information system output memory media can get be infected during installation);
- e-mail files that are sent or received check-up on viruses;
- using write-protected memory media when copying files to own hard drive; this will prevent the infecting of memory media and subsequent infection of other information systems;
- verification the memory media before files from it loading;
- permanent usage of the resident part of antiviral program which monitors all suspicious action during operation of information system.

## 4 Estimation of Software Reliability According the Results Of Adjusting and Normal Operation

In the processes of software debugging, normal and research operation, it becomes achievable to use statistical data about the detected and corrected errors in order to refine the system design reliability assessments. For this purpose, we assume to use reliability models [5-9] containing parameters, point estimations of which are obtained at the software commissioning and operation results processing. These models differ in their assumptions about the dependence of the intensity of errors emerging during the time of adjustment and operation. Some of those models contain specific requirements for the software modules internal structures.

### 4.1 Schumann's exponential model

This model is based on the following assumptions:
- the total number of commands in the program of machine language is invariable,

- at the beginning of tests, the number of errors is equal to some constant value, at length of corrections, the number of errors becomes smaller, and at course of program correcting new mistakes are not made,
- program failure rate is proportional to the number of remaining errors.

Regarding the structure of the program module there made the following assumptions:

- module contains only one cycle operator in which are resided operators for information input, assignment operators and operators of controls in advance conditional transfer,
- nested loops are absent, but there can be present $k$ parallel paths, if we have the $k - 1$ controls conditional transfer operator.

At these assumptions met, the probability of faultless operation is given by the formula:

$$R(t,\tau) = exp(-C\varepsilon_r(\tau)t) = e^{-t/T},$$

$$\varepsilon_r(\tau) = \frac{E_0}{I} - \varepsilon_B(\tau),$$

$$T = \frac{1}{\left(\tilde{N}\left(\frac{E_0}{I} - \varepsilon_B(\tau)\right)\right)}$$

(18)

where:

$E_0$ – number of errors at the beginning of adjustment,

$I$ – the number of machine instructions in the module,

$\varepsilon_B(\tau)$, $\varepsilon_r(\tau)$ – number of corrected and left errors per one command,

$T$ – mean time between failures,

$\tau$ – time of adjustment,

$C$ – coefficient of proportionality.

To assess the $E_0$ and $C$ there are used the results of adjustments. Assume that among the total number of the system test programs runs the $r$ is number of successful runs, while $n - r$ – the number of runs that were interrupted by errors. Then the total time of $n$ runs, intensity of errors and operating time per an error can be found by:

$$H = \sum_{i=1}^{r} T_i + \sum_{i=1}^{n-r} t_i, \quad \lambda = \frac{n-r}{H}, \quad T = \frac{1}{\lambda} = \frac{H}{n-r}$$

(19)

If $H = \tau_1$ and $H = \tau_2$, we may find:

$$\hat{\lambda}_1 = \frac{n_1 - r_1}{H_1}, \quad \hat{\lambda}_2 = \frac{n_2 - r_2}{H_2}, \quad \hat{T}_1 = \frac{1}{\hat{\lambda}_1}, \quad \hat{T}_2 = \frac{1}{\hat{\lambda}_2}$$

(20)

where $\hat{T}_1$ and $\hat{T}_2$ – test time for one error. Substituting here (18) and solving the system of equations, we obtain parameters for the model estimations:

$$\hat{E}_0 = \frac{I}{\gamma - 1}(\gamma \varepsilon_B(\tau_1) - \varepsilon_B(\tau_2)),$$

$$\hat{C} = \frac{1}{\left(\hat{T}_1\left(\frac{\hat{E}_0}{I} - \varepsilon_B(\tau_1)\right)\right)}, \tag{21}$$

$$\gamma = \frac{\hat{T}_1}{\hat{T}_2}$$

To compute the estimations we need, according the results of the adjustment, to learn the parameters $\hat{T}_1$, $\hat{T}_2$, $\varepsilon_B(\tau_1)$ and $\varepsilon_B(\tau_2)$.

Some generalization of results (19) – (21) look as following. Let $T_1$ and $T_2$ are times of system operation that correspond the adjustment time $\tau_1$ and $\tau_2$; $n_1$ and $n_2$ – number of errors detected in the periods $\tau_1$ and $\tau_2$. Then

$$\frac{T_1}{n_1} = \frac{1}{\left(C\left(\frac{E_0}{I} - \varepsilon_B(\tau_1)\right)\right)},$$

$$\frac{T_2}{n_2} = \frac{1}{\left(C\left(\frac{E_0}{I} - \varepsilon_B(\tau_2)\right)\right)}.$$

Hence:

$$\hat{E}_0 = \frac{I}{\gamma - 1}(\gamma \varepsilon_B(\tau_1) - \varepsilon_B(\tau_2)),$$

$$\hat{C} = \frac{n_1/T_1}{\left(\frac{\hat{E}_0}{I} - \varepsilon_B(\tau_1)\right)}, \tag{22}$$

$$\gamma = \frac{T_1}{n_1} \bigg/ \frac{T_2}{n_2}$$

If $T_1$ and $T_2$ – solely the total time of adjustment, then $\hat{T}_1 = T_1/n_1$, $\hat{T}_2 = T_2/n_2$, and formula (22) coincides with (21).

If during the adjustment course it is made $k$ tests at intervals $(0, \tau_1)$, $(0, \tau_2), \ldots (0, \tau_k)$ where $\tau_1 < \tau_2 < \ldots < \tau_k$, then to determine the maximum likelihood estimation is used equation:

$$\hat{C} = \sum_{j=1}^{k} n_j \bigg/ \left(\frac{\hat{E}_0}{I} - \varepsilon_B(\tau_j)\right) H_j,$$

$$\hat{C} = \left\{\sum_{j=1}^{k} n_j \bigg/ \left(\frac{\hat{E}_0}{I} - \varepsilon_B(\tau_j)\right)\right\} \sum_{j=1}^{k} H_j \qquad (23)$$

where:

$n_j$ – the number of runs of $j$-st test that ended with failures,

$H_j$ – time which spent on the execution of successful and unsuccessful runs of the $j$ test.

At $k = 2$ the (23) reduces to the previous case and its solution gives the result (22).

Asymptotic values of the estimations variance (for large values $n_j$) are determined according the expressions

$$\mathbf{D}\hat{C} = 1 \bigg/ \left\{\sum_{j=1}^{k} n_j / C^2 - \left(\sum_{j=1}^{k} H_j\right)^2 \bigg/ \sum_{j=1}^{k} n_j \bigg/ \left(\frac{E_0}{I} - \varepsilon_B(\tau_j)\right)^2 \right\},$$

$$\mathbf{D}E_0 = 1 \bigg/ \left\{\sum_{j=1}^{k} n_j \bigg/ \left(\frac{E_0}{I} - \varepsilon_B(\tau_j)\right)^2 - C^2 \left(\sum_{j=1}^{k} H_j\right)^2 \bigg/ \sum_{j=1}^{k} n_j\right\},$$

where $C \cong \hat{C}$, $E_0 \cong \hat{E}_0$.

The estimations correlation coefficient:

$$\rho(\hat{C}, \hat{E}_0) \cong \left\{\sum_{j=1}^{k} n_j \bigg/ \left(\frac{E_0}{I} - \varepsilon_B(\tau_j)\right)\right\} \bigg/ \left\{\sum_{j=1}^{k} n_j \sum_{j=1}^{k} n_j \bigg/ \left(\frac{E_0}{I} - \varepsilon_B(\tau_j)\right)^2\right\}^{0,5}.$$

Asymptotic value of variance and correlation coefficient are used in order to determine the confidence intervals of values $E_0$ and $C$ based on the Gaussian distribution.

Quite several studies indicate that the most appropriate for Schumann model is an exponential model of number of errors changing, along the adjustment time length changing:

$$\varepsilon_B(\tau) = \frac{E_0}{I}\left(1 - e^{-\tau/\tau_0}\right),$$

where $E_0$ and $\tau_0$ are determined empirically. Then:

$$R(t,\tau) = exp(-CE_0/Ie^{-I/I_0}t).$$

Mean time to failure increases exponentially with installation duration time increasing:

$$T = I\Big/CE_0 e^{\frac{\tau}{\tau_0}}.$$

### 4.2 Jelinsky-Morandi exponential model

This model is a particular case of Schumann model. According to this model, the errors emerging intensity is proportional to the number of residual errors:

$$\lambda(\Delta t_i) = K_{JM}(E_0 - i + 1),$$

where:

$K_{JM}$ – coefficient of proportionality,

$\Delta t_i$ – interval between the $i$ and $(i-1)$ errors.

Reliability of failure proof operation then is:

$$R(t) = exp(-\lambda(\Delta t)) = exp(-K_{JM}(E_0 - i + 1)), \ t_{i-1} < t < t_i \quad (24)$$

At $K_{JM} = C/I$ and $\varepsilon_B(\tau) = (i-1)/I$ formula (24) coincides with (18). In order to obtain maximally likelihood estimation for the parameters $E_0$ and $K_{JM}$ at sequential observation of $k$ errors in the time moments $t_1, t_2, ..., t_k$, we need to solve the system of equations:

$$\sum_{i=1}^{k}(\hat{E}_0 - i + 1)^{-1} = k\Big/(\hat{E}_0 - i + 1), \ \hat{K}_{JM} = \frac{k}{A}\Big/(E_0 - \theta \cdot k + 1) \quad (25)$$

$$\hat{K}_{JM} = \frac{k}{A}\Big/(E_0 - \theta \cdot k + 1), \ A = \sum_{i=1}^{k} t_i, \ B = \sum_{i=1}^{k} i t_i$$

Asymptotic estimations of variance and correlation coefficient (at large $k$) are determined using the formulas:

$$\mathbf{D}\hat{E}_0 \cong \frac{k}{kS_2 - A^2C^2}, \ \mathbf{D}\hat{K}_{JM} \cong \frac{S_2 K_{JM}^2}{kS_2 - A^2 K_{JM}^2},$$

$$\rho(\hat{K}_{JM}, \hat{E}_0) \cong \frac{AK_{JM}}{(kS_2)^{0,5}}, \ S_2 = \sum_{i=1}^{k}(E_0 - i + 1).$$

In order to obtain numerical values of these variables the $E_0$ and $K_{JM}$ must be replaced throughout with their estimations.

### 4.3 Weibull's exponential model

Model is given by a set of relations:

$$\lambda(t) = m\lambda^m t^{m-1}, \quad R(t) = e^{-(\lambda t)m}, \quad T = \frac{1}{\lambda}\Gamma\left(1+\frac{1}{m}\right).$$

The advantage of this model is that it contains an additional, in comparison with the exponential model, parameter $m$. Selecting the values of two parameters: the $m$ – shape parameter and $\lambda$ – scale parameter, one can get more precise correspondence with the experimental data. The values $m$ are selected from a range of $0 < m < 1$.

Parameter estimations are obtained by using the method of moments. For the shape parameter $m$, the values are found from the solution of equation:

$$\Gamma\left(1+\frac{2}{m}\right)\Big/\Gamma^2\left(1+\frac{1}{m}\right) = \frac{s^2}{\bar{t}^2}, \quad \bar{t} = \frac{1}{k}\sum_{i=1}^{k} t_i, \quad s^2 = \frac{1}{k}\sum_{i=1}^{k}(t_i - \bar{t})^2,$$

where $\Gamma(x)$ – gamma function.

For the scale parameter $\lambda$, its rating is determined according the formula

$$\hat{\lambda} = \Gamma\left(1+\frac{1}{\hat{m}}\right)\Big/\bar{t}.$$

### 4.4 Structural model by Nelson

For the reliability index is taken probability $R(n)$ of failsafe executions of $n$ program runs. For $j$ run, the probabilities of failure are as following:

$$Q_j = \sum_{i=1}^{N} p_{ji} y_i,$$

where:
  $y_i$ – an indicator of failure at $i$ set of data,
  $p_{ji}$ – the probability of $i$ set at the $j$ run.

Therefore:

$$R(n) = \prod_{j=1}^{n}(1-Q_j) = exp\left(\sum_{j=1}^{n} ln(1-Q_j)\right).$$

If the $\Delta t_j$ – time of $j$ run, the failure rate is then:

$$\lambda(t_j) = \frac{-ln(1-Q_j)}{\Delta t_j},$$

$$R(n) = exp\left(\sum_{j=1}^{n} \lambda(t_j)\Delta t_j\right), \qquad (26)$$

$$t_j = \sum_{i=1}^{j} t_i$$

Practical use of formula (26) is complicated as a result of a plurality of inputs and many hardly estimated model parameters. In practice, software reliability is assessed according the results of test trials which cover relatively small region of initial data area.

For a simplified estimation is proposed formula:

$$R(N) = \frac{1}{N}\sum_{i=1}^{N} E_i(n_i)W_i,$$

$$\sum_{i=1}^{N} W_i = N,$$

where $N$ – number of runs,

$n_i$ – number of errors at $i$ run,

$E_i$ – indicator of absence of errors at $i$ run.

To reduce the problem dimension, the multitudes of input sets are split into disjointed subsets $G_j$, to each of which corresponds a certain path $L_j, j = 1..n$. If $L_j$ has errors, then at the test performance along the subset. $G_j$ will emerge a refusal. Subsequently the probability of correct performing of single test is

$$R(1) = 1 - \sum_{j=1}^{n} p_j \varepsilon_j,$$

$$p_j = \sum_{i \in G_j} p_{ij}, \varepsilon_j < 1.$$

At this approach, to find an assessment of reliability using the structural model is difficult, since the error in $L_j$ appears not at every set from the $G_j$, but only at some of them. In addition, there is no method for the $\varepsilon_j$ estimation based on the results of programs testing.

It should be noted, that for this model, at present, has not yet been found a sufficiently reasoned justification for its implementation.

## Conclusions

Thus, in article we were analyzed in detail some methods for assessing software reliability. Many approaches have been proposed to ensure the reliability of programs, including organizational development methods, various technologies, and technological software tools, which obviously require significant resources. However, the absence of universally accepted criteria for reliability does not allow us to answer the question of how much more reliable the software becomes in compliance with the proposed procedures and technologies and to what extent the costs are justified. Thus, the priority of the task of assessing reliability should be higher than the priority of the task of ensuring it, which is not observed.

The issue of ensuring the reliability of programs is considered more important than the question of its evaluation. The situation looks paradoxical: it is obvious that before improving some characteristic, one should learn to measure it, and at least it is necessary to have a unit of measure. The main reason for this situation is rooted in the fact that the source of unreliability of the programs is the errors contained in them, and if there are no errors, then the program is reliable. Essentially, all measures to ensure the reliability of programs are aimed at minimizing (if not eliminating at all) errors in development and at the earliest possible time to identify and eliminate them after the program is made. It should be noted that error-free programs, of course, exist, but modern software systems are too large and almost inevitably contain errors. Although this circumstance has been noted by many authors and is known to any practical programmer, there seems to be a psychological barrier that does not allow recognizing the fact of errors in software as an unavoidable reality: since there is no exact criterion for determining the maximum size of an error-free program, there is always hope that in this particular software system they are gone.

The problem of choosing the unit of measurement of the reliability of a computer program cannot be solved within the framework of the industrial approach, which currently occupies an increasingly dominant position in programming.

Despite the obvious relevance, the issue of evaluating the reliability of software does not attract proper attention. At the same time, even a superficial analysis of the problem from a probability-theoretical point of view allows us to identify some patterns in this paper.